\documentclass[prb,twocolumn,floatfix,amsfonts]{revtex4-1}
\usepackage{graphicx,graphics,color,epsfig}
\usepackage{bm}
\usepackage{amsmath}
\usepackage{amssymb}
\usepackage{epstopdf}
\usepackage{epsfig}
\usepackage{graphicx}

\usepackage{longtable}
\usepackage{rotating}
\usepackage{multirow}

\begin{document}
\title{Phase-Engineered Non-degenerate Sliding Ferroelectricity Enables Tunable Photovoltaics in Monolayer Janus In$_{2}$S$_{2}$Se}

\author{Yixuan Li\textit{$^{a}$}; Qiang Wang\textit{$^{a}$}$^{\ast}$; Keying Han\textit{$^{a}$}; Yitong Liang\textit{$^{a}$}; Kai Kong\textit{$^{a}$};Yan Liang\textit{$^{b}$}; Thomas Frauenheim\textit{$^{c}$}; Xingshuai Lv\textit{$^{d}$}; Defeng Guo\textit{$^{a}$}; Bin Wang\textit{$^{e}$}$^{\ast}$}
\address{$^a$ ~State Key Laboratory of Metastable Materials Science and Technology and Key Laboratory for Microstructural Material Physics of Hebei Province, School of Science, Yanshan University, Qinhuangdao 066004, People's Republic of China. E-mail: qiangwang@ysu.edu.cn}
\address{$^b$ ~College of Physics and Optoelectronic Engineering, Faculty of Information Science and Engineering, Ocean University of China, Songling Road 238, Qingdao 266100, People's Republic of China.}
\address{$^c$ ~School of Science, Constructor University, 28759 Bremen, Germany; Computational Science and Applied Research Institute (CSAR), Shenzhen518110, P. R. China; Beijing Computational Science Research Center (CSRC), Beijing 100193, P. R. China. E-mail: thomas.frauenheim@bccms.uni-bremen.de}
\address{$^d$ ~School of Science, Shandong University, 250100 Jinan, P. R. China; Beijing Computational Science Research Center (CSRC), Beijing 100193, P. R. China. E-mail: thomas.frauenheim@bccms.uni-bremen.de}
\address{$^e$ ~Shenzhen Key Laboratory of Advanced Thin Films and Applications, College of Physics and Optoelectronic Engineering, Shenzhen University, Shenzhen, 518060, People's Republic of China. E-mail: binwang@szu.edu.cn}

\begin{abstract}

Two-dimensional sliding ferroelectrics, with their enhanced efficiencies of charge separation and tunability, constitute promising platforms for next-generation photovoltaic devices. However, recent systems predominantly exhibit dual degenerate polarization states with weak intensity, hindering the optimal manipulations of photovoltaic effects through sliding ferroelectricity. Here, we address this limitation by introducing two strengthened and distinct non-degenerate sliding ferroelectric phases ($WZ'$ and $ZB'$) in Janus $In_{2}S_{2}Se$, which can be achieved by Se-to-S substitution in monolayer $In_{2}Se_{3}$. First‐principles calculations validate the experimental synthesis of this structure and its capability for reversible phase transitions triggered by atomic layer sliding, and series of superior photovoltaic performances are demonstrated in such unique Janus $In_{2}S_{2}Se$, accompany with a detailed analysis of how non-degenerate sliding ferroelectricity modulates distinct photovoltaic characteristics. The $WZ'$ to $ZB'$ transition can increase the carrier mobility and tunes the band gap into a more moderate and indirect-to-direct character, yielding a marked red-shift and enhancement of photocurrent peak in infrared spectrum. Conversely, the $WZ'$ phase, benefiting from enhanced polarization, delivers superior photoelectric conversion efficiency in the visible light region. This work establishes a phase-engineered framework of how non-degenerate sliding ferroelectricity orchestrates distinct photovoltaic behaviors, and the intrinsic physical correlations may offer novel perspectives for next designing and regulating innovative photovoltaic devices.

\end{abstract}

\maketitle





\section{Introduction}

Two-dimensional excitonic solar cells (2D XSCs), due to its intrinsic potential for low-cost, eco-friendly, and highly efficient photo-electric conversion surpass the conventional bulk ones, is considering as ideal candidates to address recent environmental and energy crises.\cite{XSCs} In these thin films, photo-generated carriers are generated upon illumination, and then separated under varying interfacial electron ionization and affinity potentials. To date, the rapid exciton recombination remains a critical bottleneck limiting the  power-conversion efficiency in 2D XSCs.\cite{SC3} One promising strategy is integrating ferroelectricity and photovoltaics into a single system,\cite{N1} in which the spontaneous out-of-plane polarization (OOP) can effectively inhibit carrier recombination, thereby improving their photoelectric conversion efficiencies.\cite{OOP,OOP1} Recent theoretical and experimental evidences have also demonstrated this assertion.\cite{N2,RSC} However, naturally occurring 2D ferroelectric materials are exceedingly rare due to the stringent symmetry requirements.\cite{Nanom} Exploring novel formation mechanisms of 2D ferroelectricity and deciphering its role in amplifying photovoltaic responses are essential to enhance the photovoltaic metrics in next-generation of 2D XSCs.

2D sliding ferroelectrics, with the OOP polarization generated by interlayer asymmetric stacking, enabling nonpolar monolayers to acquire switchable dipoles through interlayer sliding, thus can vastly broadens the pool of 2D ferroelectrics.\cite{SC1} Such concept is initially proposed theoretically in 2017\cite{Wu}, and has been experimentally implemented since 2021.\cite{SC1,SC2} Owing to the robust structure, easy experimental operation, high Curie temperatures and fatigue-free switching,\cite{AM3} making them attractive for memory and actuator applications.\cite{SC4} Besides, the built-in polarization fields in these materials can drive superior photovoltaic effects, enabling enhanced charge separation,\cite{N3} intrinsic type-II band alignments,\cite{NanoL2} above–band-gap photovoltages\cite{SC5} and enhanced photoelectric conversion efficiencies over the limition of Shockley-Queisser\cite{N4} in 2D devices. Despite these advances, significant challenges remain in realizing ideal 2D sliding ferroelectric photovoltaic systems: One is the insufficient polarization magnitudes resulting from weak interlayer vdW coupling,\cite{Wu2} which impedes the efficient separation of photo-generated carriers; The other refers to the presence of dual degenerate polarization states,\cite{Wu2,NC1} impossible to explore the intrinsic correlations between ferroelectricity and photovoltaic effects. Further efforts should be focus on the materials design and scalable assembly techniques to unlock the full potential of sliding ferroelectrics in photovoltaics. 

Monolayer $\alpha-In_{2}Se_{3}$, featured by a corrugated quintuple-layer honeycomb structure composed of alternating Se–In–Se–In–Se atomic layers, is an ideal representative that integrates ferroelectricity and photovoltaics.\cite{SC6,NC2} Since the built-in asymmetry breaks its OOP inversion symmetry, enabling robust spontaneous OOP polarization in this emerging monolayer, which is still switchable via atomic layer sliding of middle Se.\cite{NanoL} Recent theoretical and experimental studies have demonstrated series its photovoltaic superiorities, including the narrow band gap ($\sim$1.45 eV),\cite{ACS1} the above–band-gap photovoltages,\cite{ACS2} and the photocurrents two orders of magnitude higher than bulk ferroelectrics.\cite{RSC2} The ultrafast and nonvolatile photocurrent hysteresis\cite{SC6} further highlights its potential for integrated optoelectronics and all-optical signal processing. More inspiringly, this special material is also confirmed to share two distinct ground states, namely $ZB'$ and $WZ'$,\cite{NC2} offering more versatile ferroelectric characteristics for photovoltaic modulation.

Janus monolayers, with the two faces bear different atomic species or chemical terminations, exhibit stronger OOP dipoles due to the intensified broken of inversion symmetry.\cite{N5} This asymmetric structure is first realized in transition‐metal dichalcogenides by selective replacement of one chalcogen face,\cite{Jan1} synthetic routes now extend to oxides,\cite{Nanom1} halides,\cite{N6} and rare‐earth compounds.\cite{RSC3} Inspired by these systems, the Janus $\alpha-In_{2}Se_{3}$ can be realized by substituting layers of Se atoms with the homotopic S. Compared to pristine $\alpha-In_{2}Se_{3}$, this Janus one ought to exhibit enhanced charge separation capabilities due to the asymmetrical bilateral atoms and stronger OOP dipole. In addition, such asymmetric structure also gives rise to two non-degenerate polarization states between $ZB'$ and $WZ'$, enable more adjustable polarization intensities and photovoltaic characters during their sliding transition. Therefore, we can wonder more controllable and superior photovoltaic performances in this newly predicted Janus monolayer, and further exploring the underlying correlation mechanisms between 2D sliding ferroelectricity and photovoltaics.

Here, the $In_{2}S_{2}Se$ is chosen because of its optimal band gap and superior carrier mobility among Janus various Janus $In_{2}Se_{3}$ monolayers.\cite{ACS3} First-principles calculations confirm its structural, dynamic and thermal stability. Compared to the pristine $\alpha-In_{2}Se_{3}$, its enhanced bilateral asymmetry yields stronger out-of-plane polarization and more efficient charge separation, driving improved photovoltaic performance. More strikingly, we demonstrate two non-degenerate sliding‐ferroelectric phases, $ZB'$ and $WZ'$, which can be reversible via interlayer sliding at experimentally accessible energy barriers. Further detailed analysis of its regulation effects on photovoltaics are conducted, where the $WZ'$ to $ZB'$ transition red-shifts and amplifies the infrared photocurrent peak, while the $WZ'$ phase delivers higher conversion efficiency under visible illumination. Together, these results establish a phase-engineered link between 2D sliding ferroelectricity and tunable photovoltaics, paving the way for next-generation 2D optoelectronic devices.

\section{Model and Numerical Method}

For the periodic system of Janus $In_{2}S_{2}Se$, the crystal relaxation, structural stabilities, electronic properties and light absorption coefficients were simulated by using the Vienna ab initio simulation package (VASP), which is based on density functional theory (DFT).\cite{DFT} The exchange correlation functional of Perdew–Burke–Ernzerhof (PBE) level was used to deal with the geometrical optimization and electronic structure self-consistent calculations within the generalized gradient approximation (GGA).\cite{GGA} To address the underestimation of band gap at PBE level, the more accurate results were calculated based on the hybrid density functional of Heyd–Scuseria–Ernzerhof (HSE06), and the Hartree–Fock exchange energy was set 25\%.\cite{HSE} To expand the electron wave function into plane waves, the projector augmented wave (PAW) approach was employed with a plane wave cut off energy of 500 eV.\cite{PAW} The convergence criteria were 0.01 $eV\AA^{-1}$ for force and $10^{-5}$ eV for energy, a k-point mesh of $15 \times 15 \times 1$ was used to sample the Brillouin zone, and a vacuum larger than 20 \AA was used to eliminate the spurious interactions perpendicular to the 2D plane. In addition, the dynamic stability for each phase was confirmed by the phonon spectrum, which was calculated by using the Nanodcal code. The thermal stability was verified by performing an ab initio molecular dynamics(MD) simulation within a $4 \times 4 \times 1$ supercell under 300 K, the evolution time was set 5 ps with a time step of 1 fs. The energy barriers during the sliding ferroelectric switching were calculated under the method of nudged elastic band (NEB), with the direction and intensity of OOP polarization were evaluated combining the approaches of voltage drop and Bader charge analysis.

The simulations of In$_{2}$S$_{2}$Se based two-probe nano-devices were performed by using the first-principles quantum transport package Nanodcal,\cite{PRB2, PRL} which is based on the combination of non-equilibrium Green’s function and DFT (NEGF-DFT).\cite{PRB2} During our calculation, the generalized gradient approximation at PBE level was used to handle the exchange-correlation potential, the wave functions were expanded by the basis sets using atomic orbitals of double-zeta polarization (DZP), the norm-conserving non-local pseudo-potential was applied to define the atomic core, and the energy convergence criterion of self-consistence was set to be less than $10^{-5}$ eV.\cite{DFT} Besides, a k-mesh grid of 64 $\times$ 1 for center and 256 $\times$ 1 for the leads were set perpendicular to the transport direction during the self-consistent and photocurrent computations.
    
\section{Numerical Results and Discussions}

\begin{figure*}[!tb]
	\includegraphics[width=18cm]{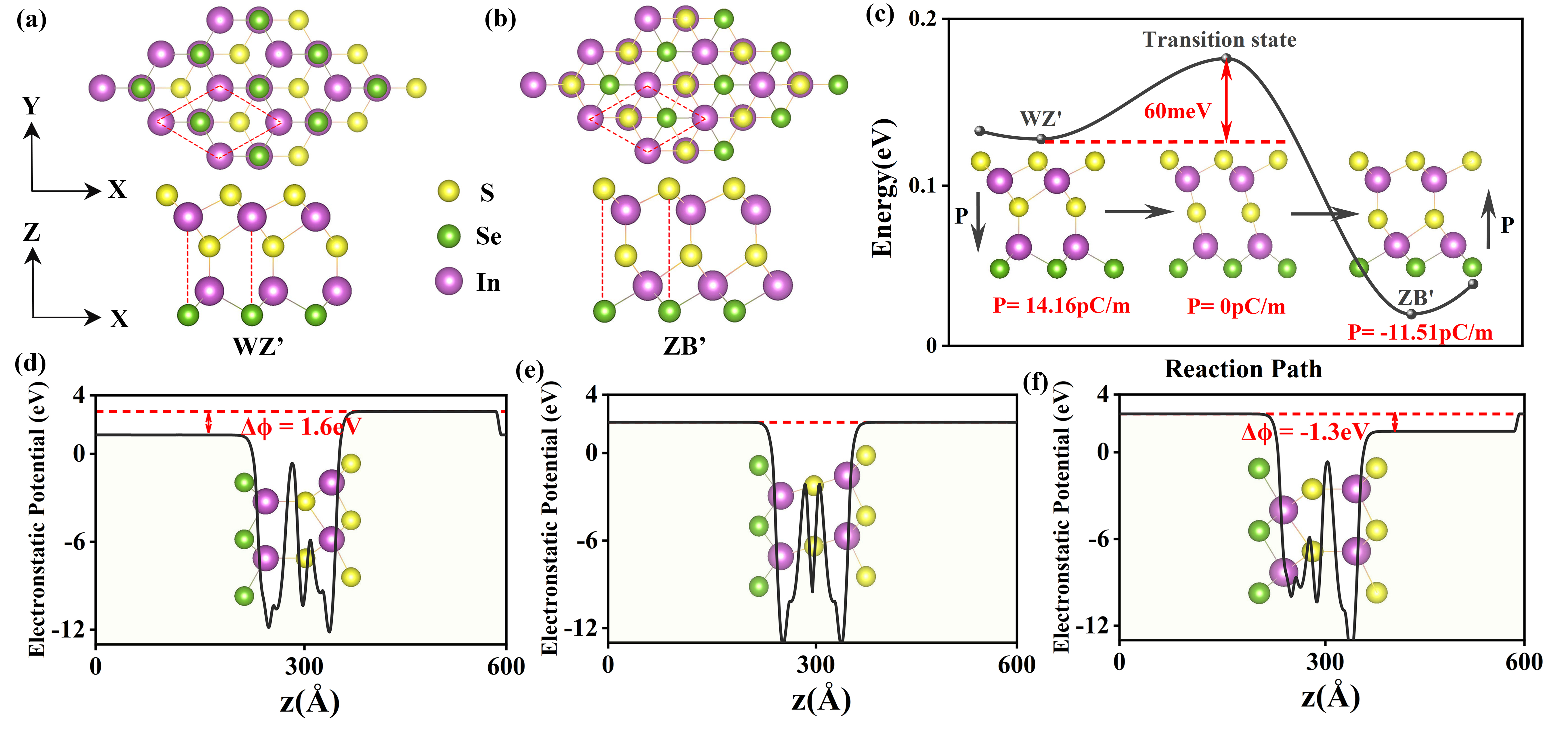}
	\centering
	\caption{(a-b)Top and side views of schematic structures of monolayer b-In$_{2}$S$_{2}$Se under $WZ'$ (a) and $ZB'$ (b) states. In each panel, the red diamond represents the size of one unit cell, with S, Se, and In atoms distinguished by yellow, green, and purple spheres, respectively. (c) Energy variation of b-In$_{2}$S$_{2}$Se along the out-of-plane ferroelectric switching pathway under NEB simulation. The three inserts depict configurations at the initial, middle, and final states, with $P$ indicating the corresponding OOP polarization strength. (d-f) Plane-averaged electrostatic potentials along the $Z$ axis for b-In$_{2}$S$_{2}$Se at different states.}
	\label{fig1}
\end{figure*}

\subsection{Structure, stability and OOP polarization of b-In$_{2}$S$_{2}$Se}

The fully optimized In$_{2}$S$_{2}Se$ structures under $WZ'$ (a) and $ZB'$ (b) states are displayed in Fig.~\ref{fig1} (a-b), where two Se atomic layers are substituted by the homotopic S. According to the different layers of atomic substitution, each state exhibits three distinct configurations: $t-In_{2}$S$_{2}Se$, $m-In_{2}$S$_{2}Se$, and $b-In_{2}$S$_{2}Se$ (Fig. S1). To obtain the most stable configuration, the binging energy for each structure is calculated using the formula: $E_{\text{b}} = \left( E_{\text{total}} - n E_{\text{In}} - m E_{\text{S}} - k E_{\text{Se}} \right) / \left( m + n + k \right)$, where $E_{total}$, $E_{In}$, $E_{S}$ and $E_{Se}$ represent the energies of monolayer material, a single In atom, S atom and  Se atom. As shown in Table s1, the $E_{b}$ of different stacking In$_{2}$S$_{2}Se$ can be reached $1.1 ~ 1.2 Jm^{2}$, larger than the typical 2D $MoS_{2}$, GaTe, and $Bi_{2}O_{2}Se$, indicating their energy feasibility. Considering the lower $E_{b}$ of $b-In_{2}$S$_{2}Se$ than others within each state, these two configurations are selected for further investigation. To examine the experimental feasibility of these two $b-In_{2}$S$_{2}Se$ monolayers, their phonon spectrum and \textit{ab initio} molecular dynamics simulations are conducted in Fig. S2. Firstly, the phonon spectra contains no virtual frequencies over the entire Brillouin zone, indicating stable minimum values of these two monolayers on potential energy, hence both of them are all dynamically stable.\cite{SZP} Additionally, to explore the thermal stability of these two monolayers, the temporal evolution of their total energy are also simulated in Fig. S2 (c-d). For each structure, the total energy fluctuates around a constant value with minimal amplitude. After 5 ps of evolutionand, only slight changes in atomic configurations occur, with no obvious geometry reconstruction or bond breaking. This comfirms that these two In$_{2}$S$_{2}Se$ monolayers maintain stability at elevated room temperature. These stability characteristics suggest robust feasibility for the experimental preparation of both $WZ'$ and $ZB'$ In$_{2}$S$_{2}Se$, and provide solid basis for their further sliding ferroelectric and photovoltaic related characters investigation. 

Inspired by the sliding ferroelectricity in monolayer In$_{2}$Se$_{3}$, we wonder whether it can be maintained in the Janus In$_{2}$S$_{2}Se$. As illustrated in Fig.~\ref{fig1} c, via specific middle and bottom atomic layers sliding, the phase transition between $WZ'$ and $ZB'$ states can be achieved in this Janus In$_{2}$S$_{2}Se$. The evolution energy barrier along the reaction path is $\sim 60 meV$, which is experimentally achievable and consisted with that of monolayer In$_{2}$Se$_{3}$ (66 meV)\cite{NC2}. Owing to the atomic asymmetry inherent to its dual-sided composition, enhanced OOP polarization intensities can be detected in such Janus In$_{2}$S$_{2}Se$, which can be reached $P_{WZ'}$ = 14.16 $pCm^{-1}$ under $WZ'$ state and $P_{ZB'}$ = -11.51 $pCm^{-1}$ under $ZB'$ state, higher than the typical monolayer $\alpha-In_{2}Se_{3}$ (10 $pCm^{-1}$),\cite{JAP2} bilayer BN (2.0 $pCm^{-1}$)\cite{SC1} and MoS$_{2}$(2.2 $pCm^{-1}$)\cite{SM} systems under identical computational conditions. 

To verify the distinct directions and intensities OOP polarization of this Janus In$_{2}$S$_{2}Se$, the electrostatic potentials under different states along Z axis are displayed in Fig.~\ref{fig1} (d-f). The potential energy decrease $\Delta \Phi$ are 1.6 eV for $WZ'$ state and -1.3 eV for $ZB'$ state, which is non-degenerated and higher than those of monolayer $\alpha-In_{2}Se_{3}$, reinforcing the stronger and non-degenerate OOP sliding ferroelectricity. Interestingly, not only such enhanced OOP polarization in Janus In$_{2}$S$_{2}Se$ may resulting in superior photovoltaic performances, the non-degenerate OOP ferroelectric states can also provide a new regulated strategy of sliding ferroelectric on photovoltaics.



\subsection{Photovoltaics related Electronic characters of b-In$_{2}$S$_{2}$Se}

\begin{figure*}[!tb]
	\includegraphics[width=17cm]{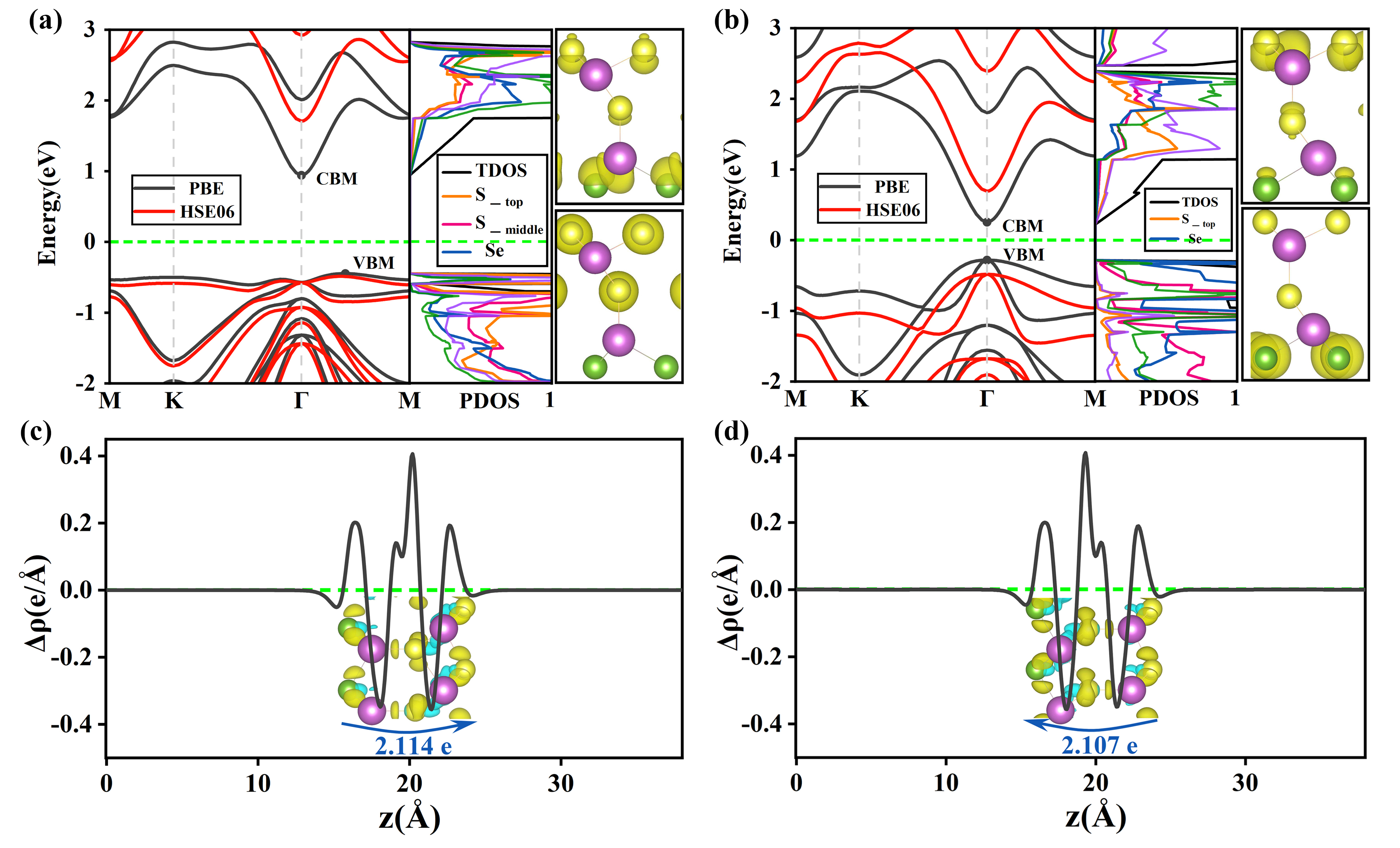}
	\centering
	\caption{(a-b) The calculated band structures (left panel), PDOSs (middle panel), and partial charge densities at CBM and VBM (right panel) for b-In$_{2}$S$_{2}$Se under $WZ'$ (a) and $ZB'$ (b) states. In each panel, the band curvess at PBE and HSE06 levels are distinguished by black and red lines, the fermi level is set to zero and marked by the green dashed line, and the isosurface value in the right panel is set at $0.015e\AA^{-3}$. (c-d) 3D isosurface and 2D integrated charge density differences along $Z$ axis of b-In$_{2}$S$_{2}$Se under $WZ'$ (c) and $ZB'$ (d) states, with isosurface value being set at $0.005e\AA^{-3}$, the charge accumulation and dissipation are indicated by green and orange balls, respectively.} 
	\label{fig2}
\end{figure*}

Next, the electric properties between $WZ'$ and $ZB'$ In$_{2}$S$_{2}$Se are investigated in-depth to explore their photovoltaic superiority. Firstly, the band structures, projected density of states (PDOS) and partial charge densities are comparatively studied in Fig.~\ref{fig2} (a-b). Obviously, with the sliding phase transition from $WZ'$ to $ZB'$, an indirect-to-direct behavior occurs accompany with a decrease of band gap. When in $WZ'$ state, the band gap are 2.19 (1.42) eV at HSE06 (PBE) level, which is too larger for visible-light absorption,\cite{SR} let along the non-beneficial of indirect band gap on the production of photogenerated charge carriers.\cite{ACIE} Inspiringly, when the phase is transfered into the $ZB'$, a moderate direct band gap of 1.17 (0.54) eV at HSE06 (PBE) level can be obtained, which perfectly fits the optimum range for excitonic solar cells (1.2-1.6 eV).\cite{MA} Another critical factor of 2D photovoltaic candidates is the capabilities of real space charge separation. As shown in the middle and right panels of Fig.~\ref{fig2} (a-b), obvious charge separation behaviors can be observed both states of the Janus In$_{2}$S$_{2}$Se. For $WZ'$ state, the conduction band minimum (CBM) is dominated by the bottom Se layer, the valence band maximum (VBM) is mainly contributed by the top and middle S layers, while under $ZB'$ state, the CBM and VBM are mainly dominated by the top S and bottom Se layers, respectively. Indeed, these distinct charge separation behaviors between the two states are driven by their intrinsic opposite OOP polarization, and further affecting their photovoltaic performances. In Fig.~\ref{fig2} (c-d), the charge density differences between the two states are displayed. Obviously opposite and significantly enhanced charge transfer across the plane can be detected of the $WZ'$ state than $ZB'$, and consisting with the more accurate Bader charge results, where the charge transfer under $WZ'$ and $ZB'$ states are 2.114 and -2.107, respectively. So far, the opposite direction and stronger intensity on OOP polarization of the $WZ'$ than $ZB'$ state are confirmed, which may further demonstrate different photovoltaic performances.

To explore the different carrier mobilities ($\mu_{2D}$) of monolayer Janus $\alpha-In_{2}Se_{3}$ between $WZ'$ and $ZB'$ states, the $\mu_{2D}$ of these two states are calculated according to the following express based on the DP theory, which has been widely used to simulate the $\mu_{2D}$ of 2D crystals:\cite{zai}
\begin{equation}
	\mu_{2D} =\frac{eh^{3}C_{2D}}{k_{B}T\lvert m^{*}_{e/h}\rvert E_{1}^{2}}
\end{equation}
where $E_{1}$ denotes the variable state, $m^{*}_{e/h}$ is the effective mass of electrons/holes, $T$ is the temperature, $k_{B}$ is the Boltzmann constant, and $C_{2D}$ indicates elastic modulus for each state crystal. The obtained results are displayed in Table s2, where $\mu_{2D}$ of Janus In$_{2}$S$_{2}$Se are obvious higher than those in monolayer $\alpha-In_{2}Se_{3}$,\cite{ACS3} suggesting superior photovoltaic performances of the former. More interestingly, since higher $\mu_{2D}$ of $ZB'$ than $WZ'$ state can be obtained, enhanced photocurrent in it based nano-device is also predicted. Over all, by expanding monolayer $\alpha-In_{2}Se_{3}$ to the Janus In$_{2}$S$_{2}$Se, we not only propose a photovoltaic system with superior performances, but can also achieve effective regulations of non-degenerate sliding ferroelectricity on photovoltaics in a single monolayer system.

\begin{figure*}[!tb]
	\includegraphics[width=17cm]{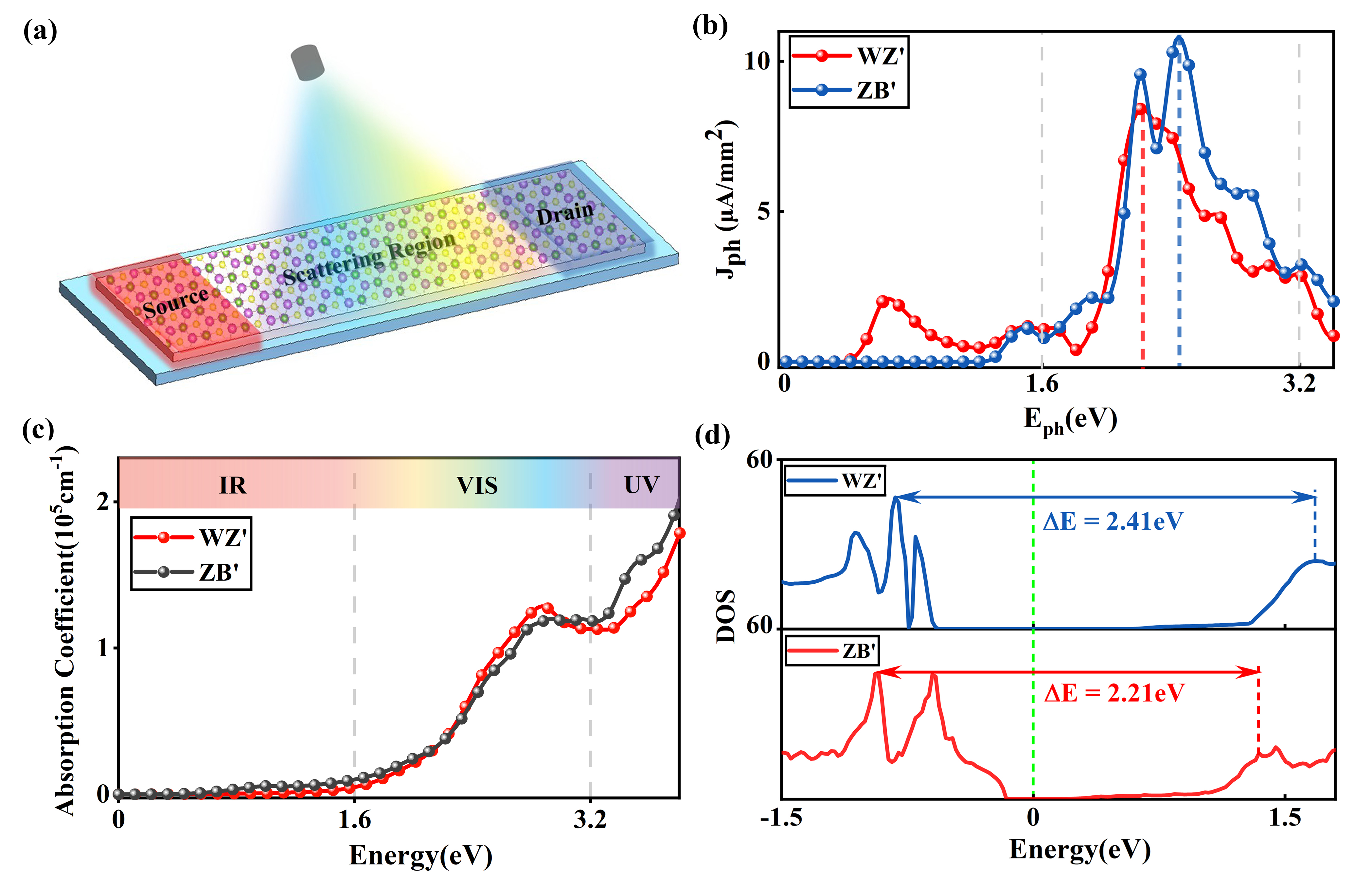}
	\centering
	\caption{(a) Schematic diagram of a two-probe PV device for the monolayer b-In$_{2}$S$_{2}$Se. (b) At $\theta$ = $0^{\circ}$, the J$_{ph}$ versus incident light energy E$_{ph}$ of the devices under $WZ'$ and $ZB'$ states. (c) Optical absorption coefficients $a(\omega)$ versus incident light energy E$_{ph}$ of $WZ'$ and $ZB'$ states. The visible light region lies between the two vertical dashed lines. (d) Density of devices states (DOS) of the $WZ'$ and $ZB'$ devices versus energy.} 
	\label{fig3} 
\end{figure*}

\subsection{Photocurrent differences between $WZ'$ and $ZB'$ b-In$_{2}$S$_{2}$Se based nano-devices}

Despite excellent photovoltaic characters of $WZ'$ and $ZB'$ In$_{2}$S$_{2}$Se have been confirmed, they are only provides qualitative predictions due to the constraints imposed by periodic boundary conditions. To further explore their actual excellence in practical devices, it is essential to account for quantum scattering under open boundary condition. Therefore, $WZ'$ and $ZB'$ In$_{2}$S$_{2}$Se based nano-devices are designed according to Fig.~\ref{fig3} (a). For each device, both leads are constructed by periodically extending the scattering region, and such simplified device model has proven to be effective and optimal in series previous theoretical and experimental research.\cite{Wq1} 

During our calculation, the entire scattering region was illuminated by linearly polarized light perpendicular to the plan. A minute bias voltage of 0.2 eV is applied between source and drain, which is far less than the band gap and only intended to drive the photocurrent. Under the first-order of Born approximation, the photocurrent ingress into the left probe can be expressed as:\cite{PRB,Xie,JAP}
\begin{equation}
	\begin{aligned}
		I_{L}^{ph}=\frac{ie}{h}\int \textbf{Tr}\left[\Gamma_{L}\{G^{<(ph)}+f_{L}(E)(G^{>(ph)}-G^{<(ph)})\}\right]dE
	\end{aligned}  \label{cur0}
\end{equation}
where $f_L$ is the Fermi distribution function of the left lead, $G^{>/<(ph)}$ represents the greater/lesser Green’s function, indicating the electron–photon interaction, and $\Gamma_L$ is the line-width function, signifying the coupling between the left lead and the central scattering region. The calculated photocurrent density $J_{L}^{ph}$ = $I_{L}^{ph}/{S}$ with polarization angle $\theta$ = $0^{\circ}$ are displayed in Fig.~\ref{fig3} (b). For both states, $J_{L}^{ph}$ start to vibrate when the incident light energy $E_{ph}$ increases to the value of band gap, subsequently increasing rapidly. Series of inspiring photovoltaic behaviors can be achieved between these two states. Firstly, significant peak values of $J_{L}^{ph}$ can be obtained under both states, with the magnitudes can be reached 10.79 $\mu A mm^{-2}$ and 8.42 $\mu A mm^{-2}$ for $WZ'$ and $ZB'$ states. Such efficient $J_{L}^{ph}$ peak values are higher than those of monolayer In$_{2}$Se$_{3}$ (3.95 $\mu A mm^{-2}$) under the same basis of calculation,\cite{ACS4} which is consistent with the above predictions in Fig.~\ref{fig2} (c-d) and can be illustrated by the enhanced OOP polarization perfectly. Besides, enlarged $J_{L}^{ph}$ of $WZ'$ than $ZB'$ state can be detected within the visible light region, Indicating its higher solar photovoltaic conversion efficiency. To delve into the intrinsic motivations, the light absorption, photogenerated carrier separation and transport characters are compared and analyzed in-depth. As shown in Figure \ref{fig3}(c), the absorption coefficients are approximately consistent throughout the entire light range of E$_{ph}$, including the infrared, visible light, and ultraviolet regions, effectively eliminating the influence of light absorption on the difference of $J_{L}^{ph}$ between the two states. In fact, such different solar photovoltaic conversions between $WZ'$ than $ZB'$ states is closely related to their distinct OOP polarization as we have discussed in Fig.~\ref{fig1}(c), where the intensified OOP polarization will drive more efficient separation of photo-generated carriers, so is the enlarged $J_{L}^{ph}$ in the visible light region. Except the magnitude of $J_{L}^{ph}$, the E$_{ph}$ positions of their peak are also different form Fig.~\ref{fig3} (b), where the peak value of $J_{L}^{ph}$ appears at E$_{ph}$= 2.41 eV for $WZ'$ state and 2.21 eV for $ZB'$ state. These different peak positions can be attributed to the distinct density of state between the two devices. As shown in Fig.~\ref{fig3} (d), the E$_{ph}$ gaps between the two initial peaks near the fermi energy are also 2.41 eV for $WZ'$ state and 2.21 eV for $ZB'$ state. The enhanced density of states can provide favorable guarantees for effective carrier transmission, and ultimately resulting in the peak of $J_{L}^{ph}$. Furthermore, a significantly enlarged and red-shifted peak value of $J_{L}^{ph}$ is observed of $ZB'$ than $WZ'$ state within the infrared region. The red-shift phenomenon can be attributed to the reduced band gap in the $ZB'$ state, which facilitates earlier oscillations of $J_{L}^{ph}$. Additionally, the augmented peak value of $J_{L}^{ph}$ is associated with the transition from an indirect to a direct band gap, as well as higher carrier mobilities in the $ZB'$ state relative to the $WZ'$ state. 

\begin{figure*}[!tb]
	\includegraphics[width=18cm]{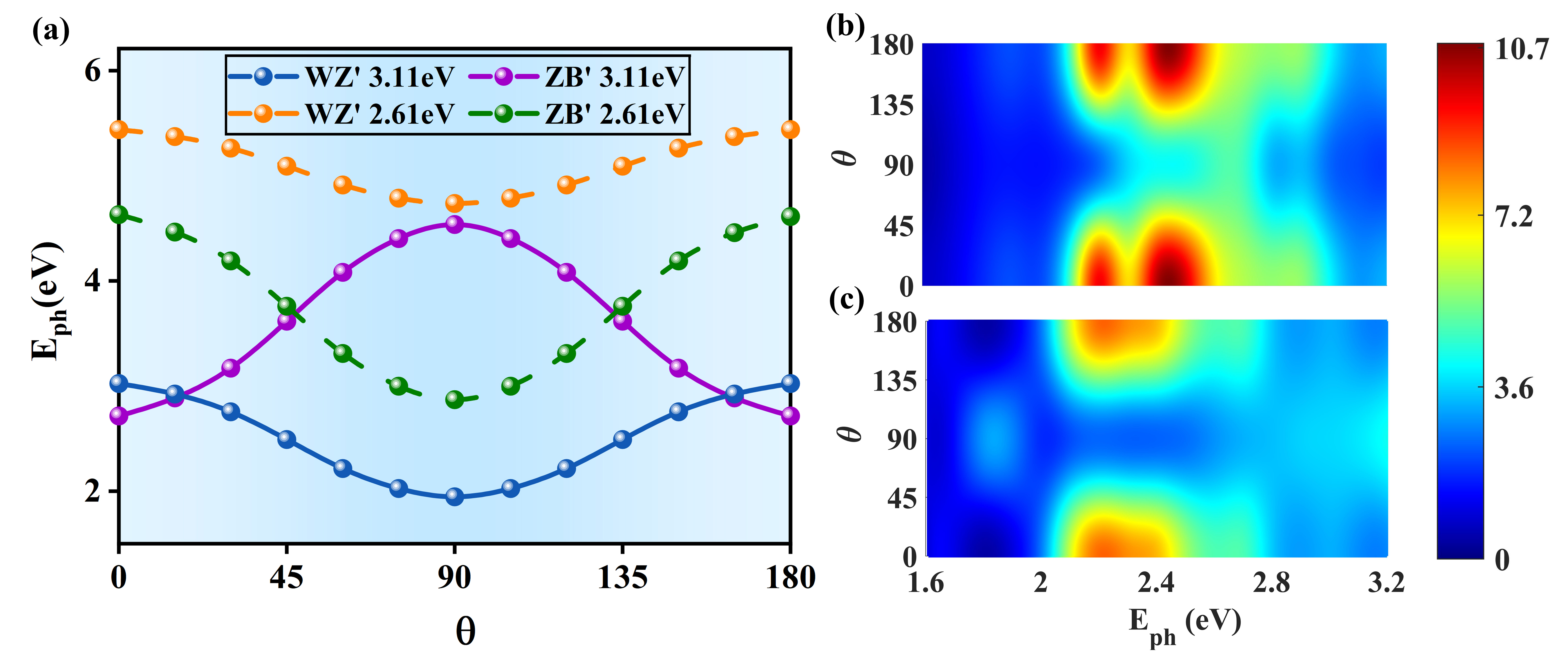}
	\centering
	\caption{(a) The $J_{ph}$ versus $\theta$ at $E_{ph} = 3.11 eV$ and $E_{ph} = 2.61 eV$ of
		the $WZ'$ and $ZB'$ state devices. Photocurrent density $J_{ph}$ versus incident light energy $E_{ph}$ and polarization angle $\theta$, for $WZ'$ (b) and $ZB'$ (c) state at the PBE level.} 
	\label{fig4}  
\end{figure*}

Moreover, distinguished evolution behaviors of $J_{L}^{ph}$ versus $\theta$ can be obtained between each state of Janus In$_{2}$S$_{2}$Se, as shown in Fig.~\ref{fig4} (a). When $E_{ph} = 3.11 eV$ reversed shapes of $J_{L}^{ph}$ curves can be observed, where $J_{L}^{ph}$ of $ZB'$ state follows the sinusoidal distribution but alongs a cosine one under the $WZ'$ state; while at $E_{ph} = 2.61 eV$, both curves obey the cosine distributions. These different evolution shapes can be reasonably illustrated by the following reformed expression of $J_{L}^{ph}$ according to Ref.[\citenum{Xie}]:
\begin{equation}
	\begin{aligned}
		J_{L}^{ph} &=\frac{ie}{h}\int\\
		&\{cos^{2}\theta\textbf{Tr}\left[\Gamma_{L}\{G_{1}^{<(ph)}+f_{L}(E)(G_{1}^{>(ph)}-G_{1}^{<(ph)})\}\right]\\
		&+sin^{2}\theta\textbf{Tr}\left[\Gamma_{L}\{G_{2}^{<(ph)}+f_{L}(E)(G_{2}^{>(ph)}-G_{2}^{<(ph)})\}\right]\\
		&+sin(2\theta)2\textbf{Tr}\left[\Gamma_{L}\{G_{3}^{<(ph)}+f_{L}(E)(G_{3}^{>(ph)}-G_{3}^{<(ph)})\}\right]\\
		&\}\times dE,
	\end{aligned}  \label{cur1}
\end{equation}
where the $J_{L}^{ph}$ is dissected into three components, which are proportional to $sin^{2}\theta$, $cos^{2}\theta$, and $sin2\theta$, respectively. Indeed, the morphological evolution of $J_{L}^{ph}$ bears a profound correlation with the corresponding crystal lattice symmetry, and can be dominated by the coefficients competition between each trigonometric functions. To elucidate further distinctions in solar photovoltaic performance between the $WZ'$ and $ZB'$ phases of In$_{2}$S$_{2}$Se, Fig.~\ref{fig4} (b-c) illustrates their $\theta$ depended $J_{L}^{ph}$ distributions entire the visible light region. In both states, the maximum peaks of $J_{L}^{ph}$ are located at $\theta$ = $0^{\circ}$, accompany with a marginal red-shift in peak position occurs in $ZB'$ state. Once again, such distinct peak positions are consisted with the $\Delta E$ between the two nearest peaks within the device's density of states in Fig.~\ref{fig3} (d). Besides, a notably elevated and broader \(J_{L}^{ph}\) peak can be observed in the $WZ'$ state, suggesting its superior efficacy in solar energy conversion compared to the $ZB'$ ones. This enhancement is ascribed to the intensified  OOP polarization in the $WZ'$ structure, which finally intensifies the separation of photo-generated carriers, a pivotal factor in photovoltaic performance. Remarkably, all above distinct excellent characters of $J_{L}^{ph}$ can be regulated via switching the two states of in Janus In$_{2}$S$_{2}$Se by sliding ferroelectricity, which is usually easier to be operated than traditional ferroelectrics in experiments.\cite{AM2,ACS5}

{\section{Discussion}}

In summary, to address the limitations of recent 2D sliding ferroelectrics in photovoltaic enhancements, we have proposed Janus $In_{2}S_{2}Se$ as a robust platform for establishing effective modulations of photovoltaics via non-degenerate sliding ferroelectricity. Due to the increased asymmetry of surface atoms, two distinct phases and stronger OOP dipoles can be realized in this Janus $In_{2}S_{2}Se$ than monolayer $\alpha-In_{2}Se_{3}$, thereby behaving superior and controllable photovoltaic related characteristics. we show that both phases of this Janus $In_{2}S_{2}Se$ are experimental feasible, and can be switchable via low‐barrier ($\sim 60 meV$) interlayer sliding. When compared to the conventional $\alpha-In_{2}Se_{3}$ monolayer, stronger OOP polarization, higher carrier mobilities, more efficient light absorption and lower exciton binding energy can be detected in Janus $In_{2}S_{2}Se$, hence enhanced $J_{L}^{ph}$ can be obtained in their based nano-devices. Besides, we have also demonstrate the phase‐dominant modulation of photovoltaic properties in this new predicted Janus $In_{2}S_{2}Se$. The $WZ'$ to $ZB'$ transition increases carrier mobilities and shifts the band gap from indirect toward direct character with more moderate size, producing a pronounced red-shift and enhanced $J_{L}^{ph}$ in the infrared spectrum. Conversely, duo to the enhanced OOP polarization in $WZ'$ phase, superior $J_{L}^{ph}$ in the visible light region is delivered during the $ZB'$ to $WZ'$ transition, indicating its more efficient solar photovoltaic conversions. Overall, by leveraging phase‐engineered sliding ferroelectricity in Janus $In_{2}S_{2}Se$, we introduce a feasible physical correlations between sliding ferroelectricity and photovoltaics, and the insight mechanisms may shed new light on conception and modulation of next ultrathin and switchable photovoltaic devices.

\section*{Conflicts of interest}

The authors declare no competing financial interest.

\section*{Acknowledgements}
This work was financially supported by grants from the Natural Science Foundation of Hebei province (Grant No. E2019203163); Innovation Capability, Improvement Project of Hebei province (Grant No. 22567605H); Cultivation Project for Basic Research and Innovation of Yanshan University of China (Grant No. 2021LGQN017); Shenzhen Natural Science Foundations (Grant No. JCYJ20190808150409413), and the Young Talents Project at Ocean University of China.






\bibliography{main} 
\bibliographystyle{elsarticle-num} 

\end{document}